\documentclass[letterpaper, 10 pt, conference]{ieeeconf}  % Comment this line out if you need a4paper

\IEEEoverridecommandlockouts                              % This command is only needed if 
\usepackage{cmap}
\usepackage{amsmath} % assumes amsmath package installed
\usepackage{amssymb}  % assumes amsmath package installed
\usepackage{graphicx}
\graphicspath{ {./images/} }

\title{\LARGE \bf
Driver Assistance Eco-driving and Transmission Control with Deep Reinforcement Learning
}

\author{Lindsey Kerbel$^{1}$, Beshah Ayalew$^{1}$, Andrej Ivanco$^{2}$, and Keith Loiselle$^{2}$% <-this % stops a space
\thanks{$^{1}$Lindsey Kerbel and Beshah Ayalew are with the Department of Automotive Engineering, Clemson University,
        Greenville, SC 29607, USA
        {\tt\small (lsutto2, beshah)@clemson.edu}}%
\thanks{$^{2}$Andrej Ivanco and Keith Loiselle are with Allison Transmission Inc., One Allison Way, Indianapolis, IN, 46222, USA
{\tt\small (andrej.ivanco, keith.loiselle)@allisontransmission.com}}%
}

\begin{document}
\setlength{\parskip}{0pt}
\bstctlcite{UpdateMaxNames}
\maketitle
\thispagestyle{empty}
\pagestyle{empty}

%%%%%%%%%%%%%%%%%%%%%%%%%%%%%%%%%%%%%%%%%%%%%%%%%%%%%%%%%%%%%%%%%%%%%%%%%%%%%%%%
\begin{abstract}
With the growing need to reduce energy consumption and greenhouse gas emissions, Eco-driving strategies provide a significant opportunity for additional fuel savings on top of other technological solutions being pursued in the transportation sector. In this paper, a model-free deep reinforcement learning (RL) control agent is proposed for active Eco-driving assistance that trades-off fuel consumption against other driver-accommodation objectives, and learns optimal traction torque and transmission shifting policies from experience. The training scheme for the proposed RL agent uses an off-policy actor-critic architecture that iteratively does policy evaluation with a multi-step return and policy improvement with the maximum posteriori policy optimization algorithm for hybrid action spaces.  The proposed Eco-driving RL agent is implemented on a commercial vehicle in car following traffic. It shows superior performance in minimizing fuel consumption compared to a baseline controller that has full knowledge of fuel-efficiency tables. 
\end{abstract}

%%%%%%%%%%%%%%%%%%%%%%%%%%%%%%%%%%%%%%%%%%%%%%%%%%%%%%%%%%%%%%%%%%%%%%%%%%%%%%%%
\section{INTRODUCTION}
Climate change concerns have prompted worldwide initiatives to reduce greenhouse gas emissions and energy consumption. For example, the Paris Climate Agreement set a goal to limit global warming by calling for substantial reductions in global  CO\textsubscript{2} emissions by 2030~\cite{ParisAgree}. The transportation sector emits 16\% of the world's greenhouse gas emissions with the majority from ground vehicles~\cite{EmSector}. To address these challenges, many technological solutions are being pursued, including light-weighting vehicles, highly efficient powertrains, and fully electric vehicles. Another approach that has received comparatively less attention is influencing driver behavior through what are known as Eco-driving strategies.

Eco-driving has been shown to reduce energy consumption anywhere between 6\% and 20\% by inducing simple changes in driving habits~\cite{ecodriv}. Reducing high accelerations, eliminating unnecessary stops, and up-shifting earlier are significant factors that reduce energy usage in all categories of vehicles. Training and monitoring drivers on Eco-driving strategies has become a critical business consideration for many fleet companies. However, long term studies have shown that the effects from initial driver training tend to fade quickly and thus reduce the benefit to 5\% just after 10 months~\cite{onboard}. To mitigate the diminishing effects of driver education, real-time feedback or continued training has been suggested~\cite{Rolim2014}. In~\cite{HeikkiUtil}, incentive programs are discussed to maintain the driver's ideal behaviors along with the many challenges of evaluating them.   

Instead of these mostly passive approaches to Eco-driving, in this paper, we consider active driver assistance control of the powertrain to achieve the goals of saving fuel while accommodating driver demand. To this end, Eco-driving, formulated as an Optimal Control Problem (OCP), allows the use of a cost function to weigh the fuel savings along with other desired outcomes.  A common OCP approach is to optimize the vehicle's velocity trajectory to minimize fuel usage. In this vein,~\cite{MaamDP} applied dynamic programming (DP) to optimize the velocity trajectory and reduce energy usage for an electric vehicle. However, DP requires the traffic, the route, and the vehicle parameters to be known a priori, which is not often satisfied. Another major group of approaches involve integrating automated Eco-driving schemes into systems such as adaptive cruise control (ACC) which, when engaged, avoids the need to rely on the driver. In~\cite{ZhaoIHV}, a 5.9\% decrease in the energy usage was observed based on implementing a strategy for ACC that maximizes regenerative braking. In~\cite{NieACCMPC}, Eco-driving was integrated into ACC with a model predictive controller (MPC) and a 12\% fuel savings was demonstrated.  

For commercial and utility vehicles in urban routes, cruise control is typically not helpful for a driver when the route requires various speeds including frequent starting and stopping.  In~\cite{Yoon}, a predictive kinetic energy management approach is proposed where radar information about preceding vehicles was used in an MPC scheme to optimize the traction torque while accommodating the driver's desired input. However, such approaches rely heavily on accurate vehicle models, not to mention means for predicting and accounting for sensing uncertainty.  For example, vehicle weight varies widely for a typical commercial vehicle, and so model-based controllers require real-time mass estimation schemes, which are very prone to error~\cite{NanMass}.

In this work, we investigate data-driven learning-based controllers that do not need models for implementation. Specifically, we formulate and demonstrate a reinforcement learning (RL)~\cite{Sutton} approach for the present application. While it is similar in objective to the optimal control approaches mentioned above, RL is fundamentally about learning the optimal policy from interactions of the RL agent with the environment without explicit needs for modeling.  Indeed, there are many emerging applications of RL to vehicle control which capitalize on the many algorithmic developments in the field in the past few years.  In~\cite{wang2018}, a RL controller is proposed for automated lane change maneuvers. Many studies have shown RL to be a near-optimal solution for controlling the vehicle velocity or acceleration in ACC systems~\cite{HuPractical}~\cite{CoopACC}~\cite{LinKin}.  A distinguishing aspect of our approach is that we seek to construct a driver-assistance Eco-driving controller for vehicular powertrains.

In this paper, we consider powertrains where both the continuous traction torque and the discrete transmission gear\footnote{In this work, we use “gear" to denote a discrete transmission state; the word “range" is also often used for the same purpose in industry.}, constitute a hybrid action space for the nonlinear system comprising of the driver-vehicle-traffic environment and the RL Agent.  Applications of model-based approaches such as MPC to such hybrid nonlinear systems lead to computationally burdensome nonlinear mixed integer programs~\cite{MPChybrid}. As a work around for the computational issues,~\cite{XuDP} applies deterministic DP to solve for optimal gear shifting on known drive cycles in an offline setting, and uses the results to train a neural network. By contrast, in this work, using the Deep RL framework, where deep neural networks (DNN) are used for both policy and value function approximation, we seek the optimal hybrid action for active driver-assistance Eco-driving.~\cite{StepGear} has sought a similar RL approach to the torque and gear control problem in the ACC setting (assuming a radar) by using a two-actor-critic architecture with a supervisor added to enforce constraints. The latter brings into question the optimality of the training as the two actors do not share the same neural networks and are optimized individually.  In our work, we create a single actor-critic hybrid action set up for driver-assistance and successfully implement the Maximum A Posteriori Policy Optimization (MPO)~\cite{neunert2020}~\cite{abdolmaleki2018} algorithm for policy improvement and the Retrace algorithm~\cite{ZhuRetrace} for policy evaluation to train our driver assist RL agent. This is likely a first application of these set of RL algorithms to this domain. In this paper, we detail our proposed RL formulation of the problem, the architecture and reward definitions, and evaluate the performance of our driver-assist RL agent. We also comment on other state of the art algorithms we have experimented with.

The rest of the paper is organized as follows. Section II outlines the problem formulation. Section III details the proposed RL framework. Section IV presents results and discussions and Section V concludes the paper.

\section{PROBLEM FORMULATION}
We formulate the problem of Eco-driving assistance control as a Markov Decision Process (MDP) as depicted in Fig. 1.  The ``environment" consists of the ego-vehicle and the driver whose states are assembled in the state vector $s=[V_{e}, A_{e}, A_{des}, n_g]$, where $V_e$ is the ego vehicle's velocity, $A_e$ is the ego vehicle's acceleration, $A_{des}$ is the driver's desired input, and $n_g$ is the transmission gear. While other states such as leading vehicle information (relative velocity and distance) can be added, for the present analysis, we treat those as encapsulated in the driver's desired acceleration $A_{des}$. In an actual vehicle, this signal is readily expressed through the driver's accelerator and brake pedal positions.  

The controller is considered as an RL agent that outputs an action vector $a=[T_{t}, u_g]$ for the desired wheel torque $T_{t}$ and a gear change command where $u_g \epsilon [-1,0,1]$ (downshift, remain in current gear, or up-shift).  $s'$ is the next state of the environment after the action is taken.  A reward $r$ is calculated from the current state and action.  Our considerations for the definition of the reward are primarily fuel consumption and driver accommodation. Further discussions about the reward will be detailed in Section III.  Other aspects of the MDP, namely the state transition dynamics including the fuel efficiency information, are a priori unknown to the RL agent. Nevertheless, we seek to learn the optimal control policies and state-action-values for the RL agent from repeated interactions of the agent with the environment. 
\begin{figure}[ht]
      \centering{
	\includegraphics[width=8.2cm]{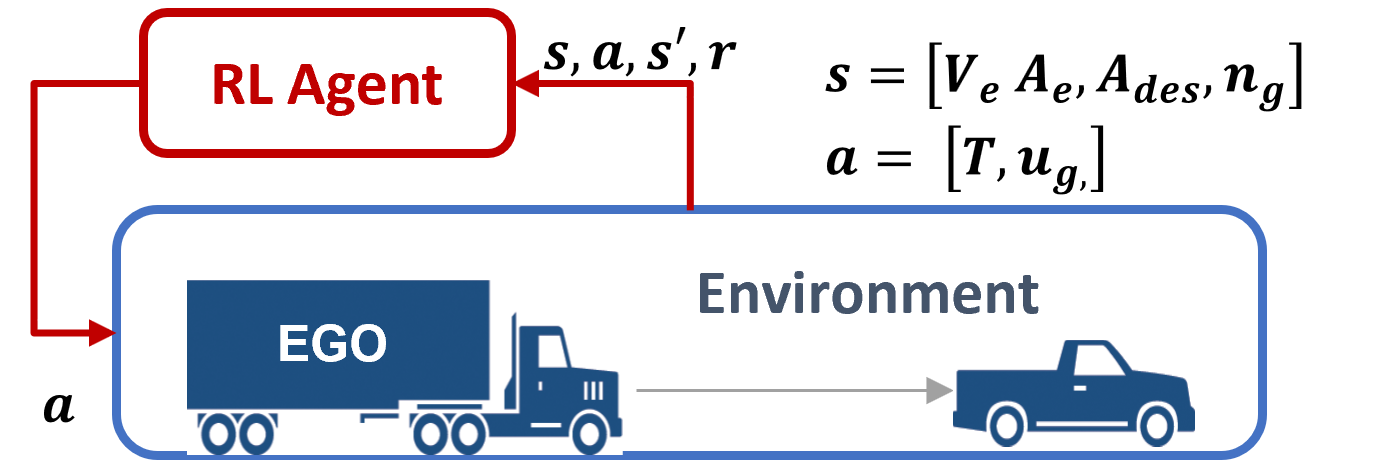}}
      \caption{RL Agent and Environment}
      \label{fig1}
 \end{figure}
\section{DEEP REINFORCEMENT LEARNING FRAMEWORK}
We approximate both the policy (actor, parameterized by $\theta$, denoted $\pi_{\theta}$) and the state-action value function (Q-value, critic, parameterized by $\phi$, denoted $Q_{\phi}$) using deep neural networks. Our overall actor-critic architecture is depicted in Fig. 2 along with the framework for the interactions with the environment shown by the solid lines.  The dotted lines represent the flow of data used to train the critic network to approximate the state-action value function and then to update the actor network parameters with the approximated Q-value. 

Since we have a hybrid action space, we train the actor network to approximate both the optimal continuous action, $a^c$, (torque, $T_t$) and the discrete action $a^d$, (gear change, $u_g$).  These actions share the same network parameters in the hidden layers, but are separated by utilizing different activation functions in the output layer. The continuous action is considered as a Guassian stochastic policy with the mean and standard deviation as its parameters. Sampled actions from this distribution are then scaled as traction torque. The discrete action is modeled as a categorical distribution through a softmax activation function with three discrete action choices.  The policy outputs requested actions to two separate powertrain actuators allowing us to consider the actions $a^c$ and $a^d$ to be independent for this work. Therefore, $\pi_{\theta}$ is factorized as~\cite{neunert2020}:  
\begin{equation} \label{eq1}
\pi_\theta(a|s) = \pi^c_\theta(a^c|s)\pi^d_\theta(a^d|s) 
  %%=  \prod_{a^i \epsilon a^c} \pi^c_\theta(a_{i}|s) \prod_{a^i \epsilon a^c} \pi^d_\theta(a_{i}|s)
\end{equation}

From the current state, the actor network computes the next actions to be sent to the vehicle powertrain. The vehicle system (in the environment) enforces operable actions consistent with the limitations of the powertrain such as engine speed $\omega_e$, engine torque $T_e$, and the available gears. It applies the available actions to the vehicle model for the transition to the next state and generates the rewards. The environment reward encompasses the fuel rate $\dot{m}_f$, the vehicle acceleration $A_e$, and gear $n_g$ that occurred as a result of the actions currently applied. The replay buffer saves the state, actions, and rewards into memory for the learning algorithm (see Fig. 2).
\begin{figure}[thpb]
      \centering{
      \includegraphics[width=8.2cm]{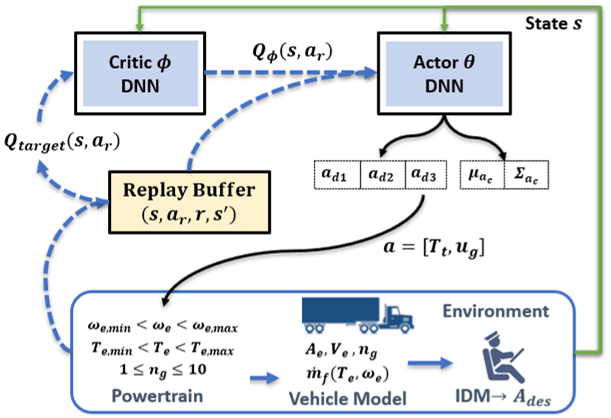}}
      \caption{Proposed Actor-Critic Architecture with Deep Neural Networks}
      \label{fig2}
 \end{figure}

We have considered state-of-the-art on-policy and off-policy algorithms to train our actor-critic set up. On-policy algorithms, such as Trust-Region Policy Optimization~\cite{schulman2017trust} and its refined version, Proximal Policy Optimization (PPO) ~\cite{schulman2017proximal} learn by using data taken from the current policy. Although simpler to tune, these on-policy methods are prone to bias and data inefficient as they learn on only the most recent data~\cite{Sutton}. With such on-policy algorithms, we would most likely need exploration and learning while driving on the road. This may mean potentially unsafe sub-optimal actions are applied to explore potential long-term rewards, not to mention the increased real-time computation burden. Despite these shortcomings, PPO was indeed implemented in a simulation setting in our work and is found to perform well in terms of convergence and showing performance benefits for the RL approach to driver assistance. However, considering the above practical considerations for eventual implementation, we sought an off-policy method. In general, off-policy algorithms are known to offer better sample efficiency, especially in high dimensional and continuous spaces. These learning algorithms typically use a batch of previous experiences saved into a replay buffer.  In this sense,  they open the option to learn from prior on-road collected data, but also explore through another facet such as a simulation model. Off-policy methods are more sample efficient as they use a behavior policy for offline exploration as discussed next.
\subsection{Policy Evaluation}
The critic network approximates a state-action-value (Q-value) for the policy.  To fit the parameters $\phi$ for the critic network, a squared loss function is minimized between the current $Q_\theta(s,a|\phi)$ and an estimated target Q-value, $Q_{target}(s_t,a_t)$.  
\begin{equation} \label{eq2}
L(\phi) =  \mathbb{E}\left[Q_\theta(s_t, a_t|\phi) - Q_{target}(s_t,a_t) \right]^2
\end{equation}
Several algorithms exists to estimate a target Q-value by taking experiences from the replay buffer.  Temporal difference is known for low variance, but high bias whereas Monte Carlo has high variance and low bias [12].  A stable and efficient method that combines the advantages of both is desirable. 

In this work, we adopted the Q-value target proposed by the Retrace algorithm~\cite{ZhuRetrace} which considers multi-step evaluation of the return as given in (3) below. It uses an importance sampling ratio $c_i$ to minimize the discrepancy from the behavioral policy $\mu(a,s)$ to the target policy $\pi(s,a)$ clipped to one to reduce variance from poor samples. Retrace uses a weighting factor $\lambda$ to exponentially decay the multi-step returns to additionally control bias and variance~\cite{ZhuRetrace}.
\begin{equation} \label{eq3}
\begin{split}
& Q^{ret}(s_t,a_t) = \displaystyle\sum_{j=t}^{t+k-1} \gamma^{j-t} \left( \prod_{i=t_1}^{j} c_i\right) \delta_j \\
& \delta = \left[r(s_j, a_j) + \gamma \mathbb{E}_\pi(Q(s_{j+1}, \cdot) - Q(s_j, a_j)\right] \\
& c_i = \lambda  \min{\left(1, \frac{\pi(a_i| s_i)}{\mu(a_i|s_i)}\right)} 
\end{split}
\end{equation}
\subsection{Policy Improvement}
A popular off-policy algorithm, Deep Deterministic Policy Gradient (DDPG)~\cite{Lillicrap2016ddpg}, which is particularly for continuous control problems, is known to be overly sensitive to hyper-parameter tuning.  In the present application, DDPG proved to be very difficult to apply.  A more recent algorithm known as MPO combines the benefits of on-policy and off-policy algorithms by drawing from a probabilistic inference perspective to optimal control~\cite{abdolmaleki2018}. We briefly summarize the key steps of this algorithm as it was implemented.

Given the value function $Q_{\theta_k}(s_t,a_t)$ for the current policy $\pi_{\theta_k}$, MPO uses two main steps to update the policy~\cite{neunert2020}~\cite{abdolmaleki2018}. The first step constructs a non-parametric improved policy $q$ from the replay buffer that maximizes $\mathbb{E}_q [Q_{\theta_k}(s_t,a_t)]$ while it stays close to the current policy. This optimization problem is posed as: 
\begin{equation} \label{eq4}
\begin{split}
& \max_{q} \mathbb{E}_{q(a|s)} [Q_{\theta_k}(s,a)] \\
& s.t.\; \mathbb{E}_{\mu(s)} \left[ KL \left( q(a|s)|| \pi_{\theta_k}(a|s) \right) \right] < \epsilon
\end{split}
\end{equation}
where $\mu(s)$ is the distribution given in the replay buffer, and the constraint is a KL divergence constraint to keep $q$ close to $\pi_{\theta_k}$. This optimization problem has the closed form solution for the sample-based distribution $q$: 
\begin{equation} \label{eq5}
q(a|s) \propto \pi_{\theta_k}(a|s) \exp \left(\frac{Q_{\theta_k}(s,a)}{\eta}\right)
\end{equation}
where $\eta$ is a temperature parameter computed by minimizing a convex dual function~\cite{abdolmaleki2018}.

The second step fits a new parametric policy $\pi_{\theta}$ to the samples from $q$ where the states are taken from the replay buffer (R). The corresponding optimization problem can be written as: 
\begin{equation} \label{eq6}
\begin{split}
& \theta_{k+1} = \arg\max_{\theta} \mathbb{E}_{s	\sim R} \left[ KL \left(q(a|s) || \pi_{\theta}(a|s) \right) \right] \\
&s.t. \; \mathbb{E}_{s \sim R} \left[ KL(\pi_{\theta_k}^c(a^C|s)||\pi_{\theta}^c(a^C|s))\right] < \epsilon_c \\
& \; \mathbb{E}_{s \sim R} \left[ \frac{1}{k} \sum_{i=1}^k KL(\pi_{\theta_k}^d(a^i|s)||\pi_{\theta}^d(a^i|s)) \right] < \epsilon_d
\end{split}
\end{equation}
In this formulation, which has been adapted for the hybrid action space factorized as in (1), the KL divergence constraint is separated into two parts, one for the discrete policy and one for the continuous policy. We further separate the KL constraints for the continuous policy into one for the mean and another for the standard deviation, thereby allowing separate tuning of each. (2) and (6) are eventually solved via a gradient-based optimization solver Adam~\cite{Kingma2015AdamAM}. The detailed derivations for the MPO algorithm can be found in~\cite{neunert2020}~\cite{abdolmaleki2018}.
\subsection{Environment Rewards}
An integral component to setting up an RL controller is the design of the reward function.  This is the premise to learning as the reward drives the state-action Q-value that evaluates the policy.  An uncertain and constrained environment increases the difficulty and the importance of motivating the desired behaviors~\cite{Dewey2014ReinforcementLA}.  Since the reward function is not directly minimized, the use of quadratic cost functions as in typical OCPs, is no longer necessary nor ideal.  Using reward functions based on the absolute values was found to have a lower steady-state error~\cite{EngelQuad}.  

The overall goal for the driver-assist controller is to balance saving energy with driver demand accommodation and reasonable control activity that doesn't compromise on comfort. The reward function that expresses this goal is given in (7).  As this is not ACC, the controller must follow the driver's desired input by minimizing the absolute error between the desired acceleration and the achieved acceleration  $(A_{des}- A_e)$. The other objectives include minimizing the controlled traction torque $T_{t}$, the fuel rate $\dot{m_f}$, and the gear shifting frequency for driver comfort.  Another driveability component is to ensure there is adequate acceleration capability available from the engine to the driver by considering a power reserve $P_r$ term. The action at time $t$ is taken on the environment at the current state $s_t$ resulting in the transition to the state at time $t+1$.  The components of the rewards are normalized with respect to their corresponding maximum values to simplify interpretations of weight selections.  
\begin{equation} \label{eq7}
\begin{split}
r(s_t, a_t) =&  -W_A\frac{|A_{des_t} - A_{e_{t+1}}|}{\Delta A_{max}} - W_{Tt}\frac{|T_{t_t}|}{T_{t,max}} \\
&- W_{fr}\frac{\dot{m}_{f_{t+1}}}{\dot{m}_{f,max}} - W_g|n_{g_{t+1}} - n_{g_t}| \\
&- W_{pr}\frac{P_{r, max_{t+1}} - P_{r_{t+1}}}{P_{r, max}}
\end{split}
\end{equation}
where $\Delta A_{max}$, ${T_{t,max}}$, and $\dot{m}_{f,max}$ are the maximum acceleration error, allowed control torque, and fuel rate. The maximum power reserve $P_{r, max}$ is a function of velocity and is computed from the maximum engine torque curve in each gear.  We make each objective a negative reward to discourage (minimize) each of the components as the MPO algorithm maximizes the total rewards.
\section{RESULTS AND DISCUSSION}
\subsection{Simulation Settings}
To demonstrate the working of the above framework, a commercial vehicle with a conventional internal combustion engine and 10-speed transmission is modeled in car-following scenarios.  The simulation of the RL controller and environment were setup in Python.  Note that the vehicle and driver model equations described below are only used to simulate the driving environment for training and evaluating the model-free driver assistance RL controller. 
\subsubsection{Vehicle and Driver Models}
The vehicle's longitudinal dynamics is modeled and simulated with (8). 
\begin{equation} \label{eq8}
\frac{dV_e}{dt}=\frac{1}{M_{eff}}\left[\frac{T_t}{r_w} +R_r (S_e)+R_a (V_e)+R_g (S_e)\right] 
\end{equation}
where $R_r(S_e)=WC_{r} \cos \psi (S_e)$, $R_a (V_e)=1/2 \rho C_d A_f V_e^2$, and $R_g (S_e)= W_e \sin \psi (S_e)$. $A_f,C_{1}$, $\rho$, and $\psi$ are the frontal area of the vehicle, the coefficient of rolling resistance, the air density, and the road grade as a function of the vehicle's position, $S_e$, respectively. The vehicle’s velocity, weight, effective mass, and wheel radius, are listed as $S_e,V_e,W_e,M_{eff}$ and $r_w$, respectively. The traction torque $T_t$ is the sum of any positive and negative torques applied to the wheels from the engine and/or the service brakes. 

In the powertrain model, the fuel rate is interpolated from a table given as a function of engine speed and torque. The engine model limits the torque based on the maximum allowable torque by the engine at the given engine speed. The negative torque is distributed first by applying the maximum engine braking allowed and the remaining torque is applied by the service brake system to the wheels.  

To simulate human-like driving behavior, various driver models have been proposed.  A well-established model for simulating typical car-following behavior is the Intelligent Driver Model (IDM).  It uses the relative distance and velocity from the preceding vehicle to calculate the desired acceleration (pedal input) of a driver (for the ego vehicle).  
\begin{equation} \label{eq9}
A_{des} = A_{max} \left[1-\left(\frac{V_e}{V_{e,des}}\right)^\delta - \left(\frac{s^*(V_e, V_r)}{S_r}\right)^2\right] 
\end{equation}
where $A_{max}$ is the maximum allowed acceleration, $V_e$ is the vehicle's current velocity, $V_{e,des}$ is the driver's desired velocity, $\delta$ is a desired acceleration exponent, $s^*$ is a function of the relative velocity $V_r$ and distance $S_r$~\cite{TreIDM}.

We also simulate a baseline powertrain controller against which we compare the performance of the RL controller. The baseline controller separates the torque and shifting control as follows. The baseline torque control consists of directly converting the driver's desired acceleration to a traction torque for the powertrain model.  The baseline shifting strategy implements a fuel optimal strategy where $u_g$ is chosen by minimizing a cost function $J_k = \dot{m}_{f,k} + q_c u_{g,k}$.  This optimization is subject to the same operating constraints imposed by the vehicle system on the RL controller. At each simulation step, the feasible gears are considered and evaluated with the cost function. While no learning or predictive attributes are added to this baseline controller, it does the gear shift optimization with full knowledge of the engine's fuel efficiency map. This advantage is deemed unavailable to the RL controller.  
\begin{table}[b]
\caption{Simulation Settings and RL Hyper-parameters}
\renewcommand{\arraystretch}{1.1}
\vspace{-0.3cm}
\label{settings}
\begin{center}
\begin{tabular}{ |cc|cc |}
\hline
\multicolumn{2}{|c|}{Vehicle Model} & \multicolumn{2}{|c|}{Hyper-parameters} \\
\hline\hline
Mass & $9070 kg$  & actor learning rate & 1e-5\\
\hline
$A_f$ & $7.71 m^2$  & critic learning rate & 1e-4\\
\hline
$C_d$ & $0.8$  & dual constraint & $0.1$\\
\hline
$r_{wheel}$ & $0.498$  & KL constraints:$\epsilon_{\mu}, \epsilon_{\sigma}, \epsilon_{d}$ & $0.1, 0.001, 0.1$\\
\hline
$C_r$ & $0.015$  & $\alpha_d, \alpha_c$ & $10$\\
\hline
$A_{max}$ & $2 (m/s)^2$  & retrace steps & $15$\\
\hline
$b$ & $3$  & $M$ action samples & $40$\\
\hline
$\delta$ & $4$  & batch size & $3072$\\
\hline
$t_h$ & $3$  & $\gamma$, $\lambda$ & $.99$, $.90$\\
\hline
\end{tabular}
\end{center}
\vspace{-4mm}
\end{table}
\subsubsection{Deep RL Network Settings}
Each of the actor and critic networks consist of 3 linear hidden layers with 256 units per layer. For the actor, a Tanh activation function is used for the mean of the continuous action, a sigmoid function for the variance, and a softmax of the discrete action choices.  The policy (actor) network weights are initialized using the Xavier Initialization method to reduce exploding or vanishing gradients~\cite{xavier}. The vehicle model and RL agent parameters are listed in Table 1.

\subsection{Analysis and Discussion}
In this work, the RL agent is trained by using a combination of the FTP and HUDDS urban drive cycles~\cite{epaRegs} as the velocity of the leading vehicle. We report results for two separate sets of reward weights labeled A and B where set B has a higher weight on the shifting frequency than set A. We refer to the corresponding RL agents as RL A and RL B. A randomization is added to the leading vehicle's velocity profile every 60 seconds to encourage generalization for changing traffic. An evaluation cycle is simulated after each training cycle where the controller uses the mean value of the Gaussian policy for the continuous action and the discrete action choice with the highest probability and has no added noise to allow for a direct comparison to the baseline. 

Fig. 3 demonstrates how the gear shifting strategy develops over multiple cycles of training. The agent learns to follow the driver's acceleration demand as it has the highest weighted reward component.  As training progresses (comparing results after 20 vs. 190 cycles), the controller learns to choose generally higher gears to minimize fuel consumption until it reaches its (near) optimal policy where the rewards no longer change significantly between training cycles. 
\begin{figure}[thpb]
\centering
      \includegraphics[width=8.2cm]{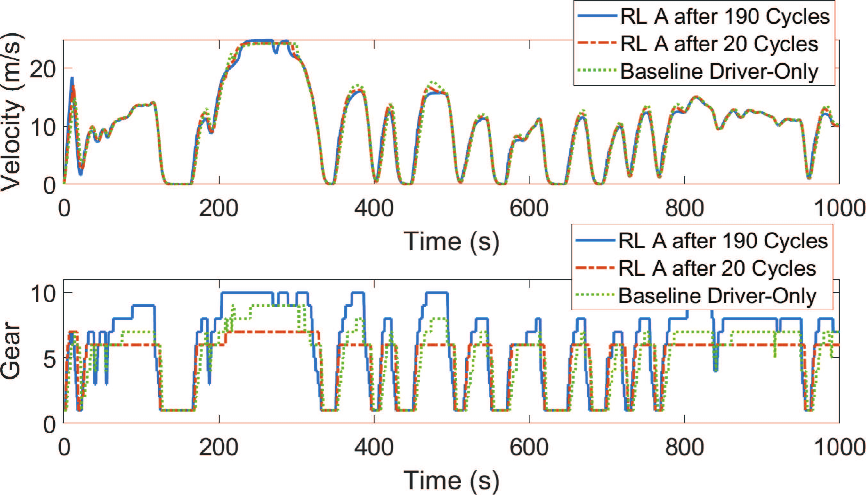}
      \caption{Evaluation of the FTP and HUDDS Cycles}
      \label{fig3}
 \end{figure}

The engine operating points for RL A and the baseline are shown in Fig. 4.  Due to the gear ratios chosen by the RL controller, the engine operates at lower speeds and higher torque where the engine efficiency is generally higher.  Although not shown here, we observed an increase in the power reserve weight leads to lower engine torque to allow for more acceleration capability. Also, in general increasing the shifting frequency penalty drives the engine operating points to higher speeds as it discourages changing gears, therefore not shifting up as early or often to reduce engine speed.  

\begin{figure}[h]
      \includegraphics[width=8.2cm]{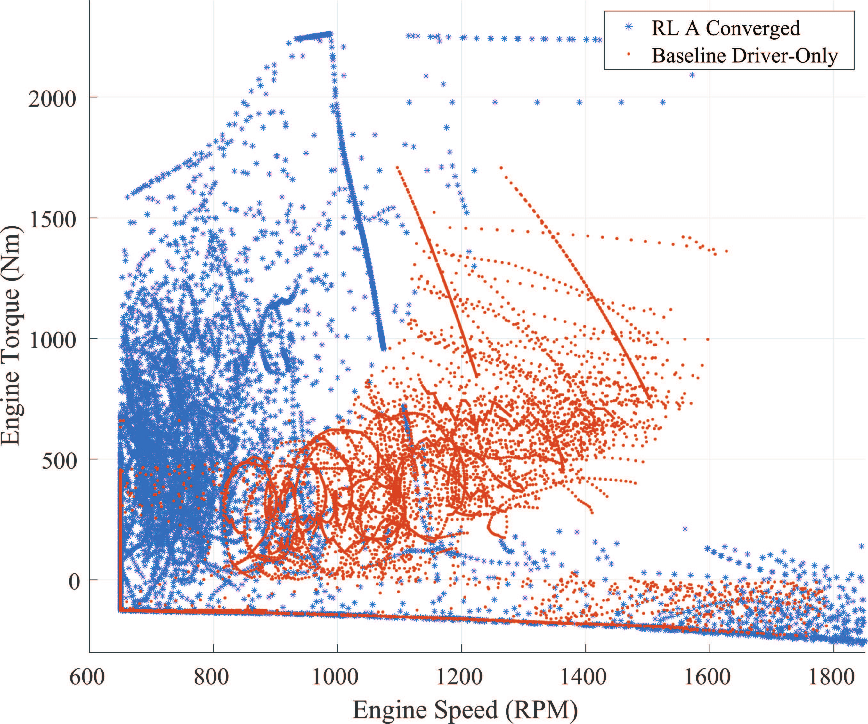}
      \centering
      \caption{Engine Operating Points During FTP/HUDDS Evaluation}
      \label{fig4}
 \end{figure}

Table 2 summarizes further evaluations of the RL agent on three drive cycles after it has reached convergence in training. The first is the combined FTP and HUDDS cycles that was used for training, but without adding randomization. To demonstrate generality of the controller, the trained RL agent was also evaluated on two different European urban drive cycles: Artemis (Urban and Road), and WLTP~\cite{EUdrivecycles} modified for heavy-duty vehicles.  Note that the RL agent was not trained on either of these additional drive cycles. 
\begin{table}[b]
\renewcommand{\arraystretch}{1.15}
\caption{RL Controller Comparison and Results}
\vspace{-0.5cm}
\label{results}
\begin{center}
\begin{tabular}{|c|c|c|c|}
\hline
\multicolumn{4}{|c|}{Cycle: FTP/HUDDS} \\ 
\hline\hline 
 & Baseline & RL A & RL B \\ 
\hline
MPG & $7.02$ & $+12.8\%$ & $+8.23\%$ \\ 
\hline
Accel RMSE $(m/s^2)$ & $0.34$ & $0.35$ & $0.13$ \\ 
\hline
$\#$ of Shifts & $466$ & $566$ & $408$ \\ 
\hline\hline
\multicolumn{4}{|c|}{Cycle: Artemis Urban} \\
\hline\hline
MPG & $5.65$ & $+11.1\%$ & $+7.69\%$ \\ 
\hline
Accel RMSE $(m/s^2)$  & $0.38$ & $0.45$ & $0.13$ \\ 
\hline
$\#$ of Shifts & $482$ & $676$ & $456$ \\ 
\hline\hline
\multicolumn{4}{|c|}{Cycle: WLTP} \\
\hline\hline
MPG & $7.3$ & $+8.26\%$ &  $+8.2\%$ \\ 
\hline
Accel RMSE $(m/s^2)$ & $0.34$& $0.30$ & $0.11$ \\ 
\hline
$\#$ of Shifts & $322$ & $389$ & $195$ \\ 
\hline
\end{tabular}
\end{center}
\vspace{-4mm}
\end{table}

The cycle fuel consumption in miles per gallon (MPG), the root mean square error (RMSE) between the desired and actual acceleration, and the number of gear changes applied throughout each drive cycle are compared to the baseline (driver-only case) for the two sets of reward weights (A and B). In this comparison, the two sets of reward weights balance the trade-off between the torque magnitude and the shifting frequency where $r_A = [W_A, W_{Tt}, W_{fr}, W_g, W_{pr} ]=[0.65, 0.095, 0.15, 0.1, 0.005]$ and $r_B =[0.65, 0.055, 0.15, 0.14, 0.005]$.  As presented in Table 2, reducing desired shifting frequency (RL B) shows less gear changes, but a lower fuel benefit.  The fuel consumption benefit was $12.8\%$ for A and $8.23\%$ for B over the baseline (driver-only) model on the drive cycle used for training (FTP/HUDDS).  The results from two additional drive cycles follow a similar trend in fuel savings and acceleration RMSE, demonstrating the robustness of the RL controller to drive cycle changes. The controller was able to track the driver's desired acceleration while maintaining a near-optimal shifting strategy even when it wasn't trained on that particular drive cycle.  Adding mass and road grade to the state variables can provide more robustness in changing environments and will be added in future work.  

It was noted that when newly initializing networks, the learning time was improved by highly penalizing known undesired behaviors.  For example, the agent initially needs to learn that to overcome the resistance forces a positive torque must be applied for the vehicle to move forward.  Other system limits, such as a gear change request that cannot be applied are currently being overridden by the vehicle system without a penalty. Finally, shift on-power technology as with an automatic transmission would reduce the importance of increased shifting frequencies somewhat in favor of higher MPG.
\section{CONCLUSIONS}
In this work, an active Eco-driving driver assistance scheme was proposed that uses a deep reinforcement learning algorithm to learn optimal torque control and transmission shifting policies. A multi-objective reward formulation was given that trades-off fuel consumption against driver-demand accommodation, shifting frequency and driveability.  The proposed RL scheme is an off-policy actor-critic scheme that uses multi-step returns for policy evaluation, and the maximum posterior policy optimization algorithm for policy improvement, customized for hybrid action spaces.  The performance of the proposed RL Eco-driving scheme is compared with that of a baseline shift and torque controller, where fuel savings of up to 12.8\% are obtained with other objectives being comparable. The observed improvements are remarkable in that the RL agent doesn't have access to efficiency tables unlike the baseline and learns better control policies in spite of that.

One major aspect of the RL framework needs further investigation: namely, the difficulty to enforce constraints during the training process. Common approaches such as reward shaping show some promise but may become too cumbersome when there are too many operating constraints in the system, some of which pertain to safety. Nevertheless, these issues need to be addressed if deep RL techniques are to receive wider acceptance in such applications.

\addtolength{\textheight}{-8.2cm}   % This command serves to balance the column lengths
\setlength{\intextsep}{-.15ex} % remove extra space above and below in-line float
\bibliographystyle{IEEEtran}
%     IEEEabrv is some IEEE template that automatically abbreviates certain journal and conference names;
%     bibliography is my actual bibliography, so I have a bibliography.bib file (does not have to be called bibliography if you name your file something else
\bibliography{IEEEabrv,bibliography} 
 \end{document}